\documentclass[]{interact}

\usepackage{hyperref}
\usepackage{makecell}

\usepackage{epstopdf}% To incorporate .eps illustrations using PDFLaTeX, etc.
\usepackage[caption=false]{subfig}% Support for small, `sub' figures and tables

\usepackage[numbers,sort&compress]{natbib}% Citation support using natbib.sty
\bibpunct[, ]{[}{]}{,}{n}{,}{,}% Citation support using natbib.sty
\theoremstyle{plain}% Theorem-like structures provided by amsthm.sty

\theoremstyle{definition}

\theoremstyle{remark}

\usepackage{color}
\newcommand{\add}[1]{\color{black} #1}

\begin{document}

\articletype{Review Article}% Specify the article type or omit as appropriate

\title{Quantifying treatment differences in confirmatory trials under non-proportional hazards}

\author{
\name{Jos\'e L. Jim\'enez 
\thanks{Email: jose\_luis.jimenez@novartis.com}}
\affil{Novartis Pharma A.G., Novartis Campus, Fabrikstrasse 2, 4056 Basel, Switzerland.}
}

\maketitle
%%REVIEWED
\begin{abstract}
Proportional hazards are a common assumption when designing confirmatory
clinical trials in oncology. With the emergence of immunotherapy and novel targeted therapies, departure from the proportional hazard assumption is not rare in nowadays clinical research. Under non-proportional hazards, the hazard ratio does not have a straightforward clinical interpretation, and the log-rank test is no longer the most powerful statistical test even though it is still valid. Nevertheless, the log-rank test and the hazard ratio are still the primary analysis tools, and traditional approaches such as sample size increase are still proposed to account for the impact of non-proportional hazards. The weighed log-rank test and the test based on the restricted mean survival time (RMST) are receiving a lot of attention as a potential alternative to the log-rank test. We conduct a simulation study comparing the performance and operating characteristics of the log-rank test, the weighted log-rank test and the test based on the RMST, including a treatment effect estimation, under different non-proportional hazards patterns. Results show that, under non-proportional hazards, the hazard ratio and weighted hazard ratio have no straightforward clinical interpretation whereas the RMST ratio can be interpreted regardless of the proportional hazards assumption. In terms of power, the RMST achieves a similar performance when compared to the log-rank test.
\end{abstract}

\begin{keywords}
log-rank; non-proportional hazards; restricted mean survival time; weighted log-rank.
\end{keywords}

\section{Introduction} %REVIEWED

Randomized controlled clinical trials are the gold standard in drug development to confirm both safety and efficacy of a new compound. The primary objective is usually to quantify the relative difference between the survival curves of the randomized treatment groups, which is commonly characterized by the hazard ratio under a proportional hazards assumption (i.e., the assumption that the ratio between the hazard function of each treatment group is constant over time). It is well known that the log-rank test is the most powerful nonparametric test under proportional hazards and thus is the most commonly used testing procedure (see \cite{schoenfeld1981asymptotic}).

Nowadays, it is not uncommon to observe a substantial departure from the proportional hazards assumption in the development of oncology drugs, for example, with the emergence of novel targeted therapies and immunotherapies. Targeted therapies point directly the oncogenic driver mutations and can lead to fast tumor regression (or disease stabilization) with fewer side effects than standard chemotherapies. However, cancer cells may develop resistance to such targeted treatment through mutation which may lead to a progression of the disease. Thus, an early separation of Kaplan-Meier curves (i.e., hazard ratio $<$ 1) followed by subsequent crossing is not uncommon for this kind of compounds. On the other hand, immunotherapies aim to boost the immune system to induce a response against the tumor. The lag between the activation of the immune cells, their proliferation, and posterior impact on the tumor is described in the literature as a delayed treatment effect. Another example where the proportional hazards assumption may not hold is when, for ethical reasons, a patient is allowed switch treatment after disease progression in a confirmatory trial. Such switching will not have an impact on progression free survival (PFS), but it may have a high impact on overall survival (OS) by diluting the treatment effect.

When the proportional hazards assumption holds, the hazard ratio captures the relative difference between the randomized treatment groups, which has clinical interpretation. However, when the underlying proportional hazards assumption is violated, the log-rank test loses power and the hazard ratio does not has a straightforward clinical interpretation as its value depends on the accrual distribution, dropout pattern and the study follow-up time, which may lead to different trial results and parameter estimates in different trials even if patients come from the same population and survival curves are identical (see \cite{schemper1992cox}). Alternative approaches to deal with non-proportional hazards patterns include the weighted log-rank test and the test based on the restricted mean survival time (RMST).

The weighed log-rank test, through the Fleming and Harrington class of weights \cite{fleming1981class}, allows to down-weight early, middle, or late events, and is proposed in the literature as a way to increase the power at the end of the trial in an intention to treat (ITT) population where treatment groups
are compared as originally randomized. Using ITT in a non-proportional hazards setting as the primary analysis may underestimate the true treatment effect although it would prevent from a type-I error rate increase (see \cite{latimer2016treatment, jimenez2020modified}). However, tuning the parameters ($\rho, \gamma$) is not straightforward since they do not have a clinical interpretation and a misspecification may cause an even larger power drop with respect to the log-rank test (see \cite{jimenez2019properties}). The Fleming and Harrington class of weights can also be used to derive a weighted hazard ratio \cite{lin2017estimation}. It can be thought as an average weighted treatment effect, although it does not have a straightforward clinical interpretation plus it has inherent ethical problems as it implies that some patients' lives are less important than others.

The RMST is a robust and clinically interpretable measure of the survival time distribution that does not require the proportional hazards assumption. Unlike the median survival time, it is estimable even under heavy censoring and has received considerable attention over the last  years (see e.g., \cite{zucker1998restricted, royston2011use, royston2013restricted, uno2014moving, uno2015alternatives}) as an alternative to estimate treatment effects under non proportional hazards. The RMST depends on the selection of cutoff (truncation) time, which needs to be pre-specified to avoid selection bias. As pointed out by \cite{huang2018comparison}, it is discussed in the literature whether the test based on the RMST may be more a sensible approach to determine superiority given the agreement in terms of statistical significance between the test based on the RMST and the log-rank test (see \cite{trinquart2016comparison}).

In a recent publication, \cite{freidlin2019methods} made a review of existing methods to test and estimate treatment effects and concluded that ``methods for accommodating non-proportional hazards such as RMST, weighted log-rank tests, and others can be useful secondary analyses because it is often difficult to have a single summary measure to accurately reflect the totality of clinical effect. However, before abandoning log-rank based primary analyses of definitive randomized clinical trials, we will need to see more convincing evidence of how these alternative methods can improve development of effective cancer therapies''. We believe that the article by \cite{freidlin2019methods} reflects to some extent the current practice in confirmatory trials where the standard log-rank test and the hazard ratio are still the primary analysis tools.

In this article, we conduct a simulation study comparing the performance and operating characteristics of the log-rank test, the weighted log-rank test and the test based on the RMST, including a treatment effect estimation, where different non-proportional patters are taken into consideration.

In Section 2, we provide a brief overview of the log-rank test, the weighted log-rank test, and the test based on the RMST, as well as how to estimate the treatment effect. In section 3 we introduce the simulation set-up we use throughout the manuscript and present the results of the simulation study. We conclude in section 4 with a discussion.

\section{Methods}
\label{ch_method}

Let $S(t) = 1-F(t)$ be the probability of survival at time $t \geq 0$, where $F(t)$ is a differentiable cumulative distribution function and $f(t)$ is its corresponding probability density function. The hazard function can then be defined as $h(t) = f(t)/S(t) = -S'(t)/S(t)$. 

If we assume that we have a control and an experimental arm, then we have a survival function $S_0(t)$ and $S_1(t)$, and a hazard function $h_0(t)$ and $h_1(t)$ for the control and experimental groups respectively. 

In a regular phase III study, we are interesting in testing the hypothesis

\begin{equation}
\label{eq_h0h1}
H_0: S_0(t) = S_1(t) \mbox{  } \forall \mbox{ } t \qquad \mbox{vs.} \qquad H_1: S_0(t) < S_1(t) \mbox{  } \exists \mbox{ } t,
\end{equation}to evaluate whether there is benefit of using the experimental treatment with respect to the control treatment, which may be the standard of care.

Let $T$ be a vector that contains the event times, Let $t_1 < \dots < t_k$ be the $k$ distinct, ordered event times. The number of patients at risk at time $t_j$  is denoted by $n_{i,j}$ with $n_j:= n_{0,j} + n_{1,j}$. Let $d_{i,j}$ denote the number of events on arm $i$ at time $t_j$ with $d_j := d_{0,j} + d_{1,j}$.

\subsection{The standard log-rank test and the hazard ratio}
\label{sc_log_rank}

The standard log-rank test statistic is then defined as
\begin{equation}
\label{logrank}
U = \sum_{j=1}^{k} \left ( d_{0,j} - d_j \frac{n_{0,j}}{n_j} \right ),
\end{equation}where the expression inside the sum describes the difference in actual and expected number of events on the control arm at each distinct time. Under $H_0$, we would have $E[U] = 0$. The variance of $U$ is given by \cite{brown1984choice} as
\begin{equation}
\label{variance_logrank}
V(U) =  \sum_{j=1}^{k} \left(  \frac{n_{0,j} n_{1,j} d_j(n_j - d_j)}{n_j^2(n_j - 1)} \right ).
\end{equation}

For large sample sizes the test statistic $Z = U/\sqrt{V(U)}$ is normally distributed with mean 0 and variance 1 under $H_0$, by the central limit theorem. For a model with proportional hazards, meaning $h_0 / h_1 = c$, where $c > 0$ is any constant, the standard log-rank test is optimal (see Schoenfeld \cite{schoenfeld1981asymptotic}) and power will increase with sample size. When the proportional hazards assumption is violated, the hazard ratio (i.e., $h_0 / h_1$) can be interpreted as an average hazard ratio (see Schemper et al. \cite{schemper2009estimation}), which is not straightforward from a clinical point of view, and the averaging depends on the overall follow-up of patients. Thus under non-proportional hazards, it is not clear to which estimand the log-rank test and the hazard ratio are associated.

\begin{figure}[h]
  \centering
  \caption{Empirical power in a toy-example with proportional hazards and delayed effects.}
  \vspace{0.5cm}
  \includegraphics[scale=0.6]{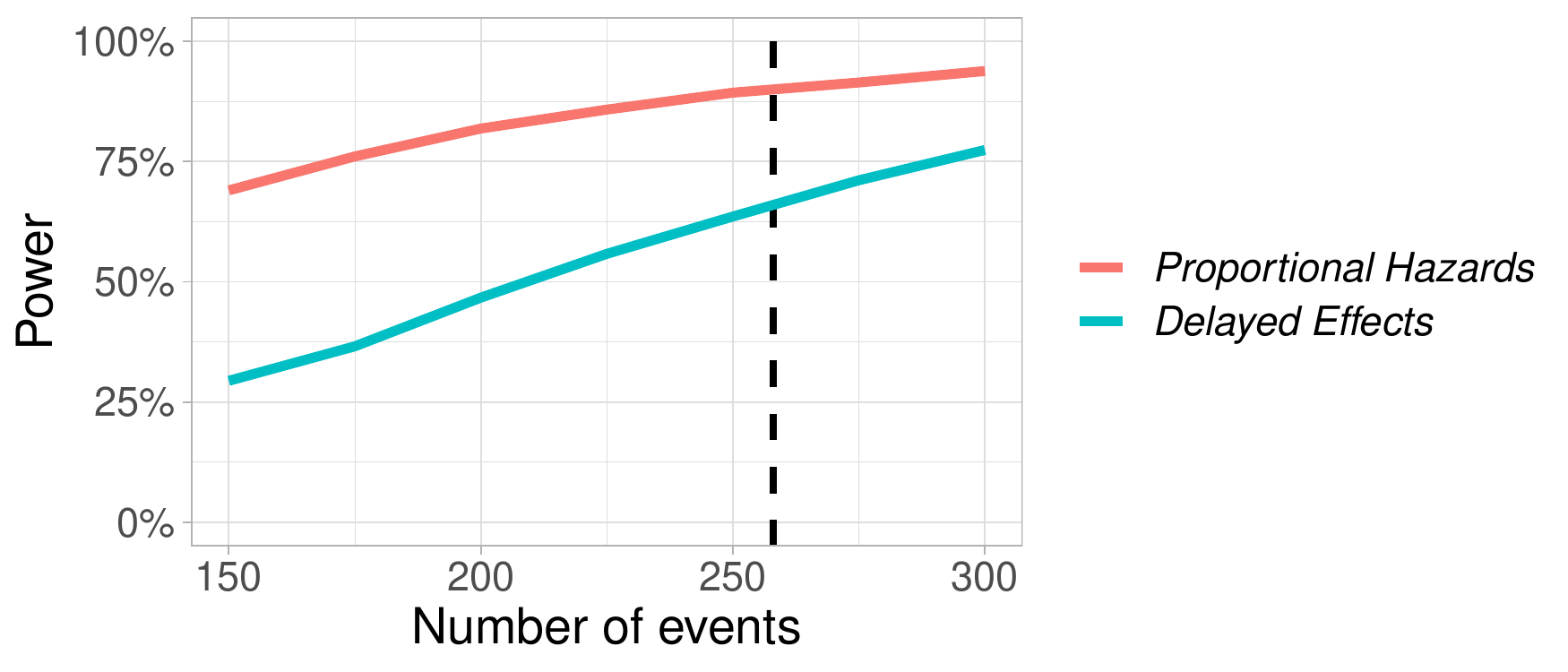}
  \label{figure1}
\end{figure}

As pointed out by Xu et al. \cite{xu2017designing}, the standard way to account for the impact of non-proportional hazards in a setting with delayed effects is either ignoring the delay or increasing the sample size but still using the standard log-rank test. In general, under the presence of delayed effects, the power increases with the sample size, although this approach may translate into unrealistically big sample sizes for middle to large delays. In Figure \ref{figure1} we present a toy-example that shows how the power behaves when defined as a function of the number of events under proportional hazards and delayed effects. This example uses the simulation set-up described in section \ref{sc_simulation_setup} where 258 events are required to achieve a 90\% power under proportional hazards. At 258 events and with a delay 2 months, we observe that power goes from 0.90 to 0.67, This translates into a relative efficiency of $\left ( \frac{\Phi^{-1}(0.975) + \Phi^{-1}(0.90)}{\Phi^{-1}(0.975) + \Phi^{-1}(0.67)}\right )^2= 1.82$, which means that we would need 82\% more patients if we would choose to use the standard log-rank test in this delayed effects example.

Therefore, accounting for the impact of the delayed effects by increasing the number of events, although theoretically possible, would not be feasible from a practical point of view given the potentially large additional sample size the trial would need to assume. On the other hand, if we choose it ignore the delay, the trial will not be sufficiently powered and most likely will not be accepted by any health authority under normal circumstances.

\subsection{Weighted log-rank tests and the weighted hazard ratio.}
\label{sc_weighted_log_rank}

An alternative to the standard log-rank test is the weighted log-rank test, defined as

\begin{equation}
\label{weighted_logrank}
U = \sum_{j=1}^{k} w_j \left ( d_{0,j} - d_j \frac{n_{0,j}}{n_j} \right ),
\end{equation}
with variance
\begin{equation}
\label{variance_weighted_logrank}
V(U) =  \sum_{j=1}^{k} w_j^2 \left(  \frac{n_{0,j} n_{1,j} d_j(n_j - d_j)}{n_j^2(n_j - 1)} \right ).
\end{equation}

The rational for this weighted test comes from the assumption that under non-proportional hazards we expect a higher treatment effect in a particular time of the study. Intuitively, by down-weighting for example late events, we may achieve higher power than the standard log-rank test in a setting with delayed effects.

The test statistic is defined as $Z = U/\sqrt{V(U)} \sim N(0,1)$ under $H_0$ \cite{fleming1981class}. By setting the weights $w_j = 1$ (or any other constant) we get the standard log-rank test.

A general class of weighted log-rank tests was introduced by Fleming-Harrington \cite{harrington1982class} $G_{\rho,\gamma}$ with a weight function of the form $w_j = (\hat{S}(t_j))^{\rho} (1 - \hat{S}(t_j))^{\gamma}$ where $\rho \geq 0, \gamma \geq 0$, and $\hat{S}(t_j)$ is the estimated pooled survival function immediately prior to time $t_j$. Several log-rank tests can be derived with different $\rho$ and $\gamma$ combinations. 

\begin{figure}[h]
  \centering
  \caption{Power for each $\rho$ and $\gamma$ combination of the Fleming and Harrington class of weights in a scenario with a median OS for the control group of 6 months, a median OS for the experimental group of 9 months, and a delay of 4 months.}
  \vspace{0.5cm}
  \includegraphics[scale=0.7]{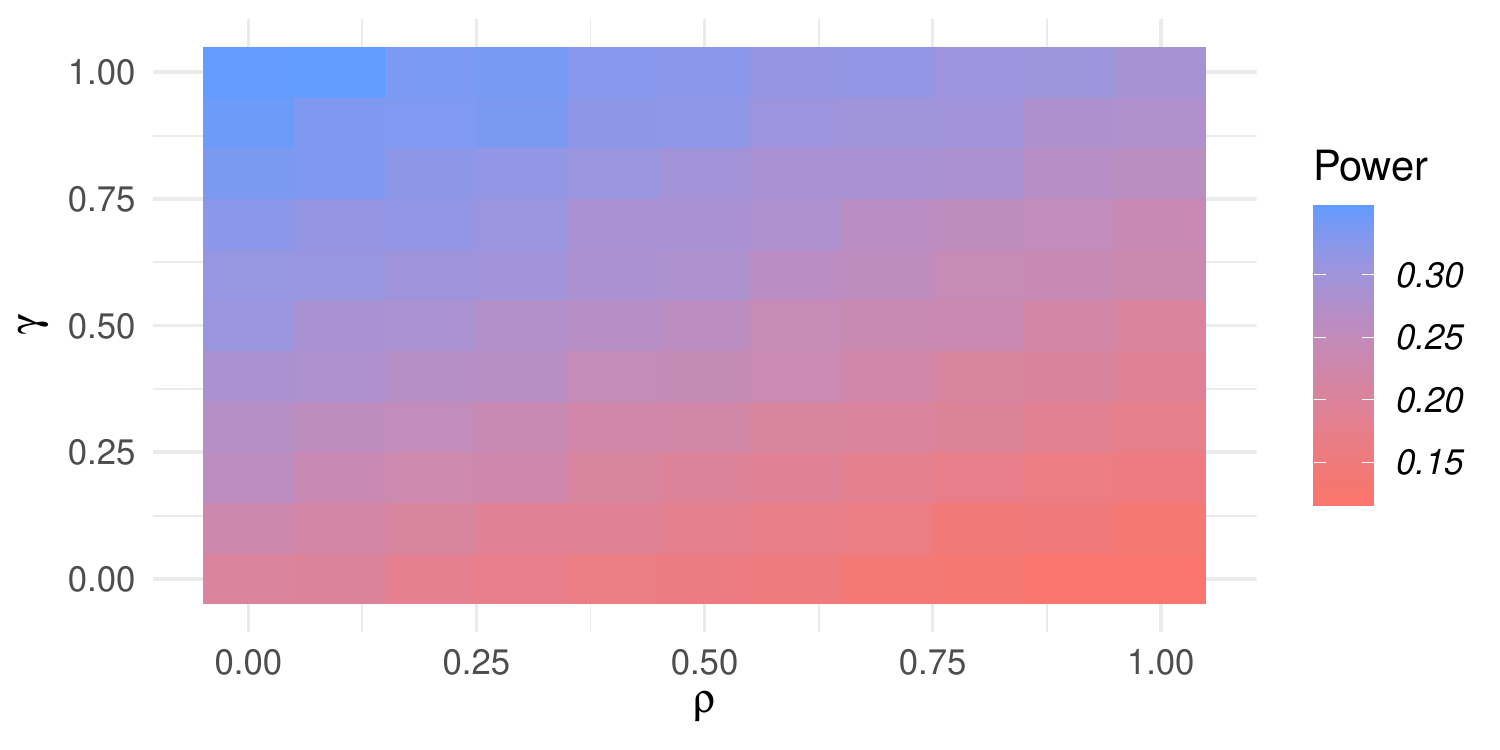}
  \label{power_wlr_example}
\end{figure}

For example, with $(\rho=1, \gamma=0) = G_{1,0}$ we obtain the Prentice-Wilcoxon test, where higher weights are assigned to early survival differences. With $G_{0,1}$ and $G_{1,1}$ we emphasize late and mid differences and with $G_{0,0}$ we obtain the standard log-rank test. In Figure \ref{power_wlr_example} we show the estimated power in a toy-example under the presence of delayed effects with all $\rho$ and $\gamma$ combinations where we see how $G_{0,1}$ is clearly the most powerful parameter combination for this type of non-proportional hazards. See \cite{jimenez2019properties} for an extensive evaluation of the use of the Fleming-Harrington class of weights under non-proportional hazards caused by delayed effects. The choice of $\rho$ and $\gamma$ requires extensive knowledge of the shape of survival curves and plays a key role that may result in loss of power of the weighted log-rank test. Therefore, due to the uncertain nature of non-proportional hazards, specification of $\rho$ and $\gamma$ is difficult. 

As a possible solution, \cite{lee1996some} proposed a versatile max-combo test, which takes the maximum value of a set of different $G_{\rho,\gamma}$, each of which is most powerful in detecting a certain pattern of non-proportional hazards. The multiple testing adjustment is conducted via a Dunnett-type parametric method.

The treatment effect linked to weighted log-rank test is known as the weighted hazard ratio (see e.g., \cite{harrington1982class,schemper1992cox,lin2017estimation}), a weighted average treatment effect where the weights are those from the associated weighted log-rank test. For instance, in a delayed effects setting, we would use $G_{0,1}$ to down-weight the early treatment effect and be more reflective of the hazard ratio in the later part of the curve. This would measure the average weighted treatment effect, which does not have a straightforward clinical interpretation plus it has inherent ethical problems as it implies that some patients’ lives are less important than others.

%%%%%%%%%%%%%%%%%%%%%%%%%%%%%%%%%%%%%%%%%%%%%%%
%%%%%%%%%%%%%%%%%%%%%%%%%%%%%%%%%%%%%%%%%%%%%%%
\subsection{The restricted mean survival time (RMST)}
\label{sc_rmst}

The RMST has been proposed as an alternative summary for the survival curve (see \cite{irwin1949standard, uno2014moving, uno2015alternatives, trinquart2016comparison}) and is defined the expected value of survival time up to a fixed time point $t^*$. In other words, the RMST is the average survival time estimate which corresponds to the area under the Kaplan-Meier curve up from the beginning of the trial until time $t^*$. Inferences about the RMST have been extensively discussed in the literature (see e.g., (\cite{karrison1987restricted, zucker1998restricted, royston2011use, tian2014predicting}). 

Moreover, it is currently under discussion whether using the ratio (or difference) of two RMST may be more sensible to determine superiority given also the agreement in terms of statistical significance between the test based on the RMST and the log-rank test (see \cite{uno2014moving, uno2015alternatives, pak2017interpretability, huang2018comparison, trinquart2016comparison}).

Let $\mu(t^*)$ be the mean of a survival function truncated at time $t^* > 0$. This corresponds to the area under the survival curve $S(t) = P(T>0)$, and thus $\mu(t^*) = \int_0^{t^*} S(t)dt$. Also let $\sigma^2(t^*) = 2 \int_{0}^{t^*} t S(t) dt - \left[ \int_0^{t^*} S(t) dt \right ]^2$. To estimate $\mu(t^*)$ in we can use the Kaplan-Meier estimator of $\hat{S}(t)$, and therefore $\hat{\mu} = \int_0^{t^*} \hat{S}(t)dt$, where $\hat{\mu}(t^*)$ approximately follows a normal distribution with variance $V(\hat{\mu}(t^*)) = \sum_{i=1}^D \left [ \int_{t_i}^{t^*} \hat{S}(t) dt \right ]^2 \frac{d_i}{Y_i(t_i - d_i)}$, where $d_i$ and $Y_i$ are the number of events and number of subjects at risk at time $t_i$ respectively.

Let the estimated difference between treatment arms in terms of RMST be defined as 

\begin{equation}
   U =  \int_{0}^{t^*} \left ( \hat{S}_1(t) - \hat{S}_0(t) \right ) dt = \hat{\mu}_1(t^*) - \hat{\mu}_0(t^*), 
\end{equation}where $\hat{S}_1(t)$ and $\hat{S}_0(t)$ are the estimated survival curves of the experimental and control groups respectively. The estimated variance term is defined as

\begin{equation}
V(U) = V(\hat{\mu}_1(t^*)) + V(\hat{\mu}_0(t^*)),
\end{equation}and the test statistic as $Z = U/\sqrt{V(U)} \sim N(0,1)$ under $H_0$.

Another alternative would be to compute $\int_0^{t^*} \hat{S}_0(t) dt / \int_0^{t^*} \hat{S}_1(t) dt$, a measurement of the relative risk similar to the hazard ratio (i.e., a ratio below 1 implies a treatment effect in favor of the experimental arm) with a variance term estimated using the delta method.

{\add One of the limitations of the RMST is that the RMST value depends on the value of $t^*$}, which is constrained to the duration of the follow-up and the censoring. This is found quite frequently in the literature (see e.g., \cite{freidlin2019methods}) as a justification for selecting the log-rank test (and hazard ratio) over a RMST-based approach. However, the log-rank test is a sequence of hypergeometric tests that requires both the number of patients at risk and number of events to be greater than zero (see \cite{fleming2011counting}). This implies that the log-rank (and by extension the hazard ratio) depends on a time-window. Moreover, as pointed out by \cite{tian2020empirical}, the maximum time-window allowed by the RMST is actually wider than the one for the log-rank test, which means that the RMST uses more data than the log-rank test. {\add This is very well illustrated in \cite{huang2020estimating} (Figure 1A), where we can see how the the time-window for RMST difference is larger than the time-window of the log-rank, indicating that RMST uses more data than log-rank.}

On this matter, \cite{zhao2016restricted} proposed to summarize the survival distribution via the RMST up to a sequence of $t^*$'s in an interval. In theory, these intervals should be pre-specified based on clinical and feasibility consideration to know what possible time window we can choose to compute the RMST estimates.

\begin{figure}[h]
  \centering
  \caption{Power as a function of the cutting time $(t^*)$ in the RMST under proportional hazards, delayed effects, crossing hazards and decreasing effects using the simulation set-up described in section \ref{sc_simulation_setup}. The thresholds where the hazard ratio changes for the delayed effects, crossing hazards and decreasing effects non proportional hazards pattern are 4, 9 and 9 months, respectively.}
  \vspace{0.25cm}
  \includegraphics[scale=0.55]{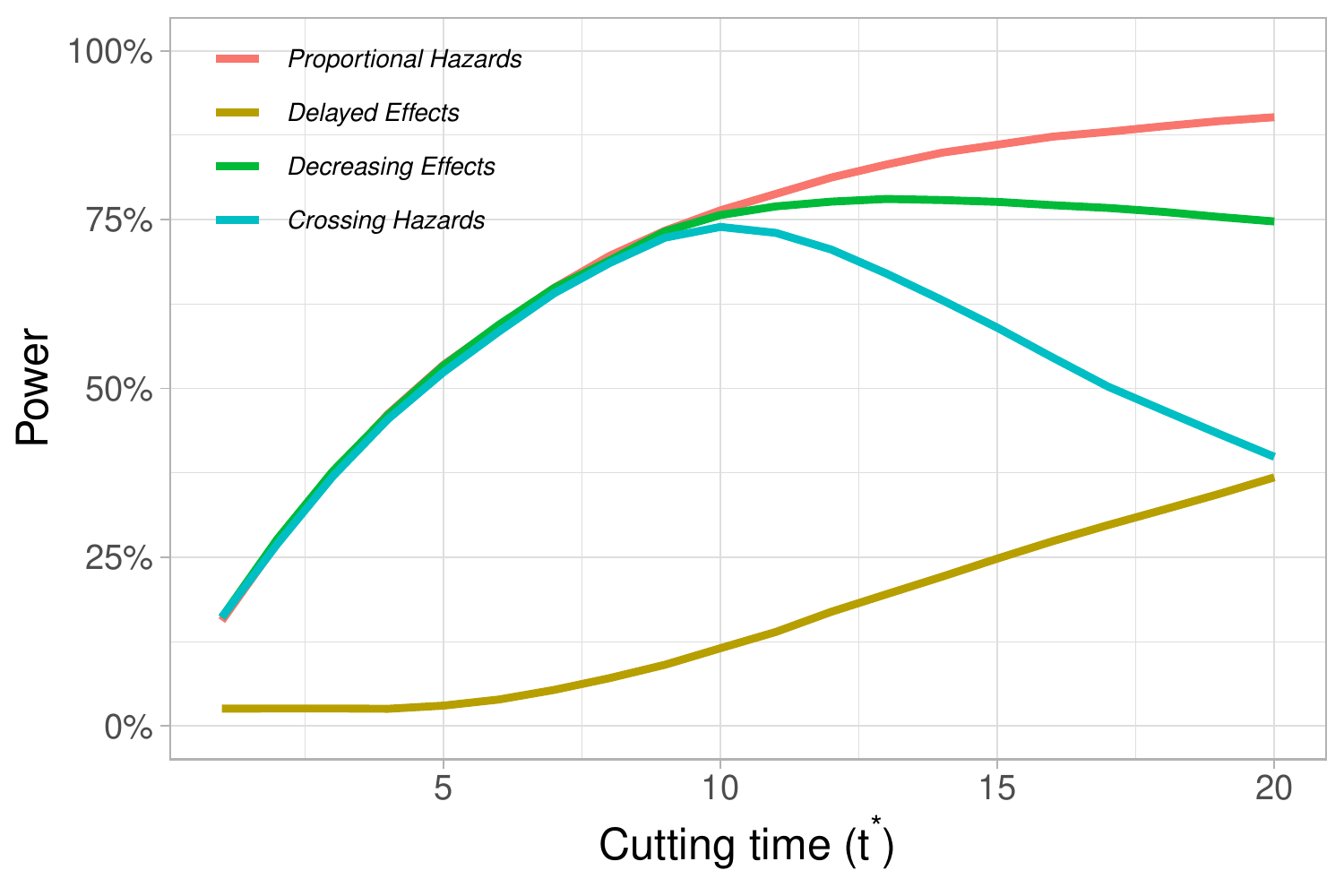}
  \label{figure3}
\end{figure}

In Figure \ref{figure3} we present the power of the test based on the RMST in the setting described in section \ref{sc_simulation_setup} as a function of the cutting time $(t^*)$ under proportional hazards and under non-proportional hazards. We can observe how, under decreasing effects and crossing hazards patterns, increasing $t^*$ causes power loss, which is consistent since true treatment effects tend to disappear over time in both settings. In the delayed effects and proportional hazards patterns, increasing $t^*$ causes a power increase. The selection of $t^*$ has been briefly discussed in several papers (see e.g., \cite{royston2011use, royston2013restricted, trinquart2016comparison}) and it is recommended that $t^*$ is pre-specified to minimize selection bias and to protect the integrity of the trial. Moreover, and as pointed out by \cite{huang2018comparison}, $t^*$ should be clinically meaningful and closer to the end of the study follow-up so that most survival outcomes are covered by the time interval $[0,t^*]$. In this article, in order to have a fair comparison with the approaches based on the log-rank and the weighted log-rank tests, $t^*$ is linked to the data and is defined as the minimum of the maximum observed (event or censored) time of each treatment group (i.e., minimax observed time).

Another possible $t^*$ could be the minimum of the maximum event time of each treatment group (i.e., minimax event time). However, as pointed out by \cite{huang2018comparison}, for delayed effect settings, $t^*$ equal to the minimax event time could lead to poor outcomes, while for settings with crossing hazards and belly-shape curves, $t^*$ equal to the minimax event time performs slightly better than the minimax observed time. 

\section{Simulated study}

\subsection{Setup}
\label{sc_simulation_setup}

In this section we describe the simulation set-up we use in this article. Following \cite{jimenez2019properties}, we employ a scenario that imitates a realistic phase III in oncology where survival data is simulated from a piece-wise exponential distribution. Under proportional hazards, we assume that the control group has a median OS of 6 months while the experimental group has a median OS of 9 months. Thus, the true hazard ratio (i.e., full effect) is equal to 0.667.

\begin{figure}[h]
  \centering
  \caption{Hazard functions from the 3 non-proportional hazard patterns used in this article assuming hazard ratio change threshold (e.g., when the effect of the immunotherapy kicks in a delayed effects setting) of 4 months.}
  \vspace{0.2cm}
  \includegraphics[scale=0.55]{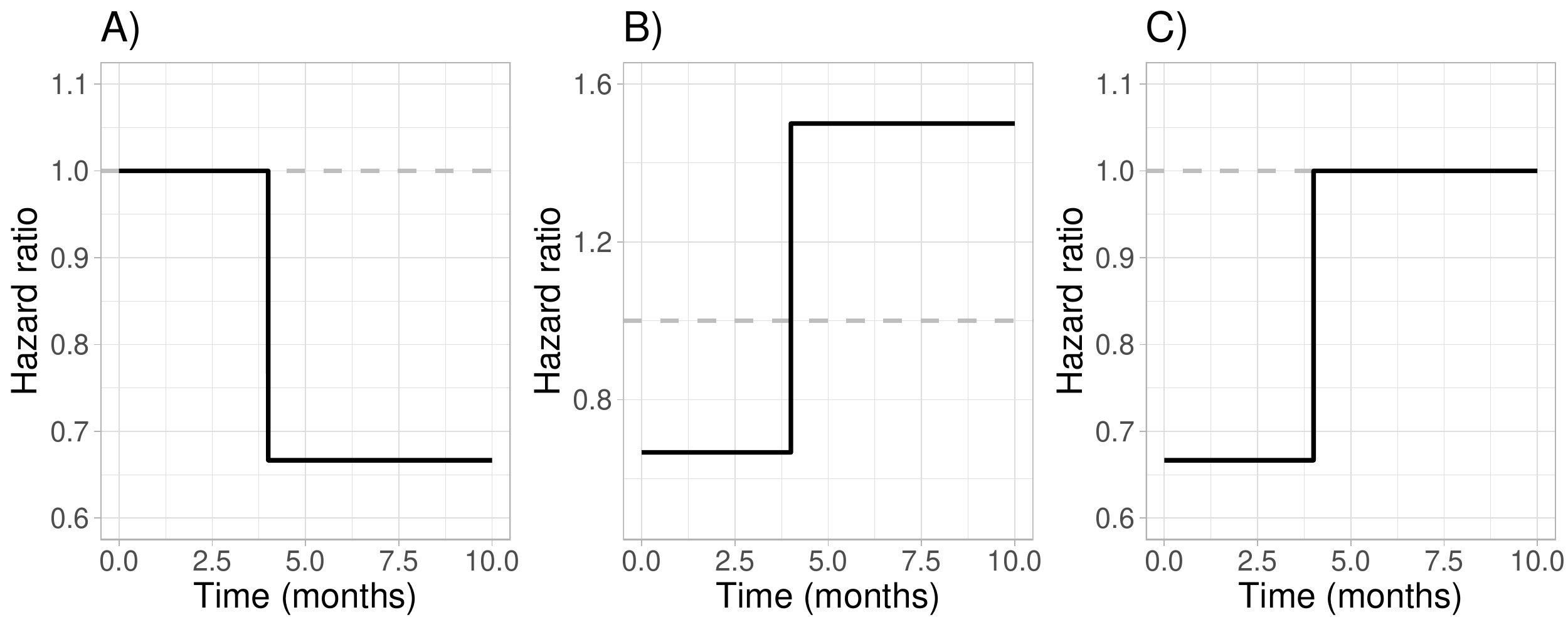}
  \label{figure4}
\end{figure}

In Figure \ref{figure4} we present the 3 non-proportional hazard patterns under study in this article where the time in which there is a change in the hazard ratio function is chosen to be equal to 4 months for illustration purposes. We establish a total study duration of 25 months, a total enrollment period of 17.5 months, randomization ratio 1:1, a power of 90\% and a one-sided level $\alpha$ of 2.5\%. We assume an enrolment of 330 patients with a minimum of 258 events to achieved the desired operating characteristics when the assumption of proportional hazards. All the results are based on $10
^4$ simulated trials implemented in \texttt{R}. In this article we assume a 22\% of censoring rate, including potential dropouts.

Under delayed effects, as shown in Figure \ref{figure4}A, the hazard ratio is assumed to be equal to 1 until the time in which there is a change in the hazard ratio function, and equal to 0.667 afterwards. Under crossing hazards, as shown in Figure \ref{figure4}B, the hazard ratio is assumed to be equal to 0.667 until the time in which there is a change in the hazard ratio function, and equal to 1.5 afterwards. Under decreasing effects (Figure \ref{figure4}C), the hazard ratio is assumed to be equal to 0.667 until the time in which there is a change in the hazard ratio function, and equal to 1 afterwards. In the simulation study, for each non-proportional hazards pattern, different threshold values are used and power is calculated with the log-rank test, the weighted log-rank test with the Fleming-Harrington class of weights $G_{0,1}, G_{1,1}$ and $G_{1,0}$ (see section \ref{sc_weighted_log_rank}) and with the test based on the RMST. Treatment effect will be quantified using the hazard ratio, the weighted hazard ratio using $G_{0,1}, G_{1,1}$ and $G_{1,0}$ and both the difference and the ratio between the RMST of each treatment group.

\subsection{Power and treatment effect}
\label{sc_power}

\subsubsection{Delayed effects} %REVIEWED

\begin{figure}[h]
  \centering
  \caption{Power (A) and treatment effect (B) estimation using the standard log-rank test, weighted log-rank test and RMST in a delayed effects setting.}
  \vspace{0.5cm}
  \includegraphics[scale=0.55]{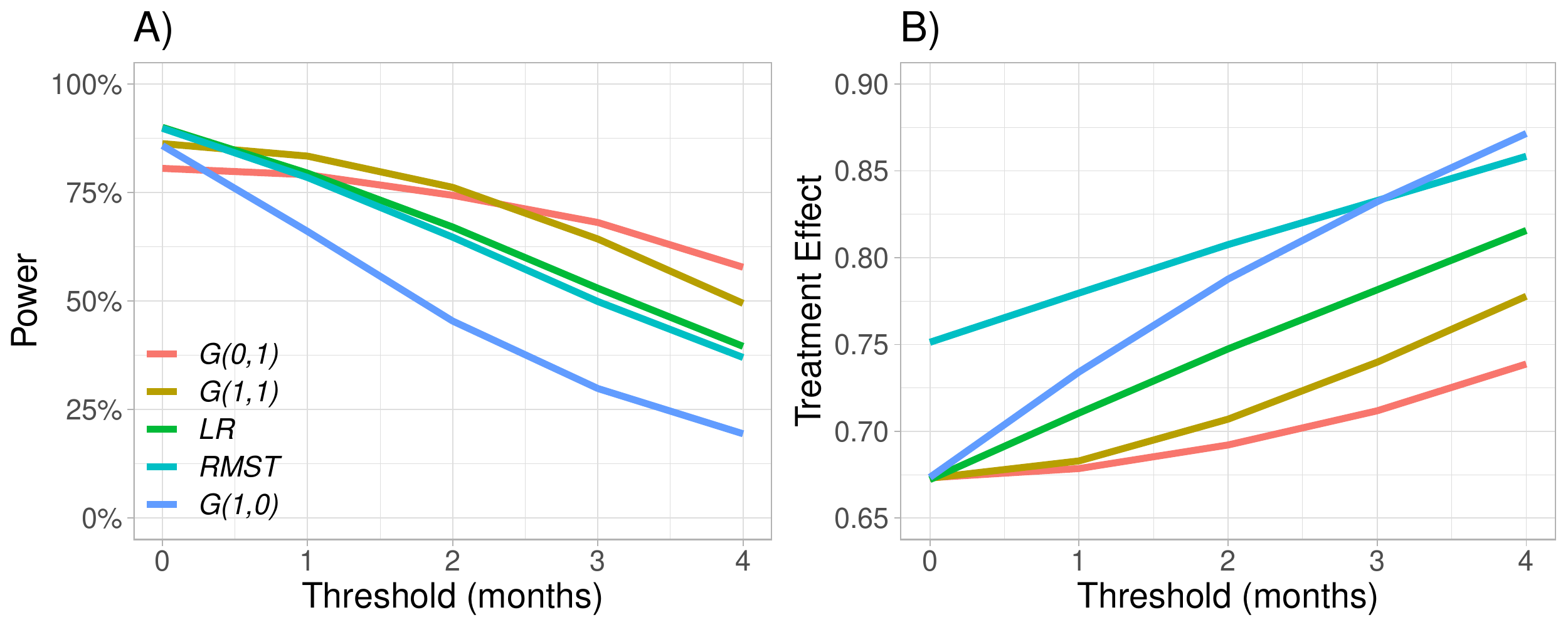}
  \label{figure5}
\end{figure}

Figure \ref{figure5}A shows the empirical power and estimated treatment effects of the scenario described in section \ref{sc_simulation_setup} under delayed effects with a threshold that indicates the change in the hazard ratio function (i.e., delay) ranging from 0 (i.e., proportional hazards) to 4 months. In terms of power, we observe a general decrease as the delay increases. We notice that, as expected, the weighted log-rank tests $G_{(0,1)}$ and $G_{(1,0)}$ are the tests that achieve the highest and lowest power overall, respectively. The standard log-rank test and the test based on the RMST achieve very similar power values regardless the delay value.

In terms of the quantification of the treatment effect we observe that, as the delay increases, the estimated treatment effect increases towards 1 as presented in Figure \ref{figure5}B. The hazard ratio, represented with the ``LR'' line, has a treatment effect of 0.67 under proportional hazards that increases up to 0.82 when the threshold that indicates the change in the hazard ratio function is equal to 4 months. As previously mentioned, 0.87 would be an average hazard ratio which does not a clear clinical interpretation.

The value obtained from the weighted hazard ratio represents the hazard ratio linked to the weighted log-rank test, which down-weights early, middle or late events conditional to the selection of $\rho$ and $\gamma$ (see section \ref{sc_weighted_log_rank}). These treatment effects estimates are represented in Figure \ref{figure5}B with the lines ``$G_{01}$'', ``$G_{10}$'' and ``$G_{11}$''. We can see how when the threshold that indicates the change in the hazard ratio function is equal to 4 months, $G_{01}$ yields an estimated treatment effect of 0.73, which is close to the full effect, whereas $G_{10}$ yields an estimated treatment effect of 0.87. Nevertheless, these results do not have a straightforward clinical interpretation even if the estimated treatment effect of $G_{01}$ seems to be closer the full effect when $S_1(t) > S_0(t)$. Moreover, it is not clear to which estimand they are linked and, as previously mentioned, it has inherent ethical problems as it implies that some patients’ lives are less important than others due to the use of $G_{\rho, \gamma}$. 

The ratio between RMST is represented in Figure \ref{figure5}B with the ``RMST'' line and regardless the value of the threshold that indicates the change in the hazard ratio function, we can always interpret it from a clinical point. We can see how under proportional hazards, the ration between RMST is equal to 0.75 under proportional hazards and 0.86 when threshold that indicates the change in the hazard ratio function is equal to 4 months. To have a better understanding of these values, in Table \ref{table1} we present all the RMST for each treatment group and for each threshold value considered in this setting. We observe that the RMST values of the control group barely change and range from 7.9 and 8 months. This is because the control group does not have any delayed effect and hence the average survival time does not change. In contrast, the RMST values of the experimental group start at 10.6 months under proportional hazards and decrease until 9.2 months when the threshold that indicates the change in the hazard ratio function is equal to 4 months. The rational for this behaviour is that if the threshold is higher, the time until we can observe the full effect will also be higher which makes the average survival time of the experimental group be closer the average survival time of the control group. Hence, it is straightforward to interpret and understand why when the threshold is equal to 4 months, the ratio between RMST is equal to 0.86, and how as the threshold increases, the ratio between average survival times will tend towards 1. In Table \ref{table1} we also provide the difference between the RMST of each group, where we see that these difference tend toward 0 as the threshold increases.

\begin{table}[h]
\centering
\caption{RMST of each treatment group in a delayed effects setting for values of the threshold that indicates the change in the hazard ratio function ranging from 0 (i.e., proportional hazards) to 4 months.}
\vspace{0.25cm}
\begin{tabular}{|c|c|c|c|c|c|}
\hline
\makecell{Threshold (months)} & 0    & 1    & 2    & 3   & 4   \\ \hline
\makecell{RMST in control group} & 8.0 & 8.0 & 7.9 & 7.9 & 7.9 \\ \hline
\makecell{RMST in experimental group} & 10.6 & 10.1 & 9.8 & 9.4 & 9.2 \\ \hline
\makecell{Difference between RMST} & 2.6 & 2.1 & 1.9 & 1.5 & 1.3 \\ \hline
\end{tabular}
\label{table1}
\end{table}

\subsubsection{Crossing hazards} %REVIEWED

\begin{figure}[h]
  \centering
  \caption{Power (A) and treatment effect (B) estimation using the standard log-rank test, weighted log-rank test and test based on the RMST in a crossing hazards setting.}
  \vspace{0.25cm}
  \includegraphics[scale=0.55]{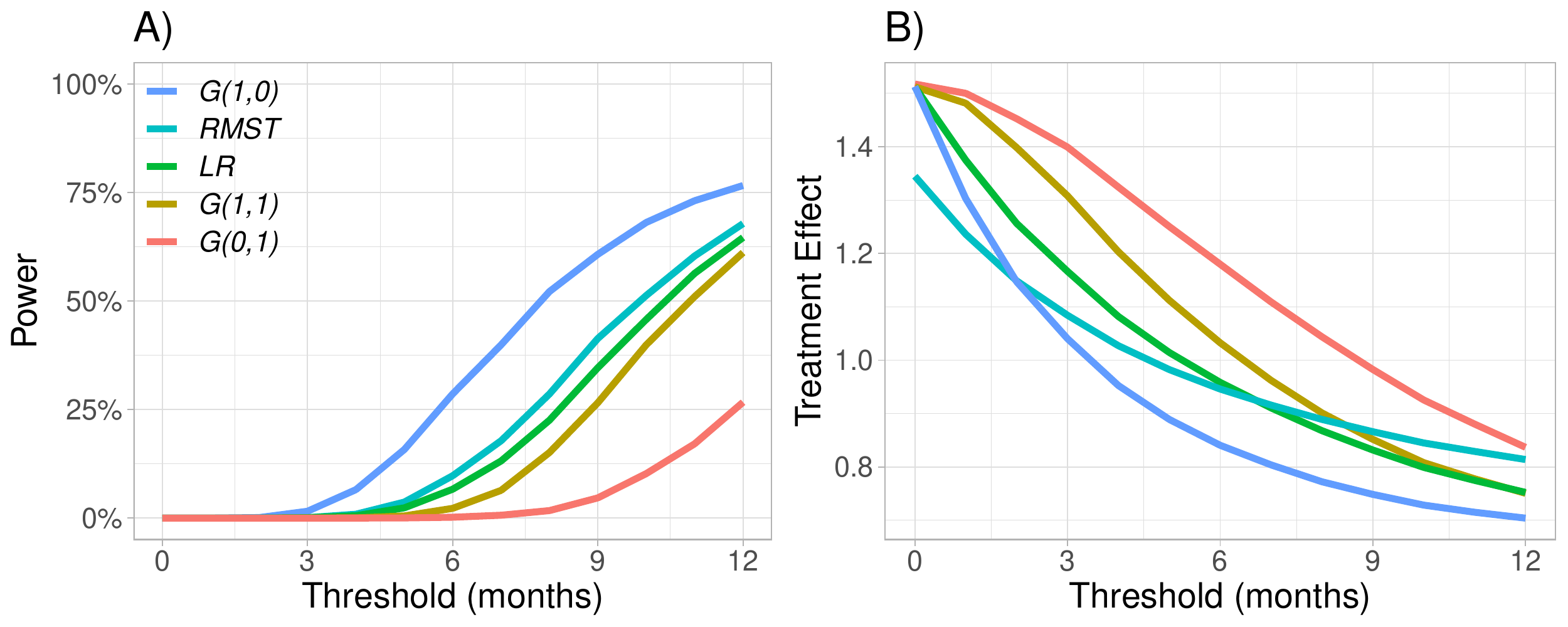}
  \label{figure6}
\end{figure}

In Figure \ref{figure6} we present the empirical power and the estimated treatment effects of the proposed scenario under crossing hazards with a threshold that indicates the change in the hazard ratio function ranging from 0 to 12 months. In terms of power, we observe a general increase as the threshold that indicates the change in the hazard ratio function increases. This behaviour is consistent since when the threshold is equal to 0, $S_0(t) > S_1(t)$ throughout the entire study and therefore we never reject $H_0$ (i.e., power is equal to 0). Then, as the the threshold increases, the time throughout the study in which $S_1(t) > S_0(t)$ is also higher, and hence it is more likely to reject $H_0$, which causes a power increase. We notice that, as expected, the weighted log-rank tests $G_{(1,0)}$ and $G_{(0,1)}$ are the tests that achieve the highest and lowest power overall, respectively. We again notice that the standard log-rank test and the test based on the RMST achieve very similar power values regardless the value of the threshold that indicates the change in the hazard ratio function.

In terms of the quantification of the treatment effect we observe that, as the threshold increases, the estimated treatment effect decreases, which is consistent with the nature of this type of non-proportional hazards since the higher the threshold the higher amount of time throughout the study in which $S_1(t) > S_0(t)$, which would translate into treatment effect estimates in favor of the experimental group (i.e., treatment effects below 1). The hazard ratio represented in Figure \ref{figure6}B with the ``LR'' line, has a treatment effect equal to 1.5 when the threshold is equal to 0, and equal to 0.75 when the threshold is equal to 12 months. However, this estimate would be an average hazard ratio which does not a clear clinical interpretation under non-proportional hazards and could be only interpretable when the threshold is equal to 0, which is an extreme case placed here for illustration purposes. In fact, 1.5 is the true hazard ratio (i.e., full effect) in this setting when $S_0(t) > S_1(t)$ as we can see in Figure \ref{figure4}B.

The weighted hazard ratio represented in Figure \ref{figure6}B with the ``$G_{01}$'', ``$G_{10}$'' and ``$G_{11}$'' lines also start yield a treatment effect estimation of 1.5 the threshold is equal to 0. However, when the threshold is relatively high, say 6 months, we can see how $G_{01}$ and $G_{10}$ yield a treatment effect 1.18 and 0.85 respectively. The reason for these differences is that $G_{01}$ down-weights early events and $G_{10}$ down-weights late events. Nevertheless, these results do not have a straightforward clinical interpretation even if the estimated treatment effect of $G_{10}$ seems to be closer the full effect when $S_1(t) > S_0(t)$. These results also highlight the impact of misspecifying the values of $\rho$ and $\gamma$ which do not have a clinical interpretation.

The ratio between RMST is represented in Figure \ref{figure6}B with the ``RMST'' line. To have a better understanding of these values, in Table \ref{table2} we present all the RMST for each treatment group and for each threshold value considered in this setting. We observe that the RMST values of the control group barely change and range from 7.4 and 7.8 months. Again, this is because the control group does not have any delayed effect and hence the average survival time does not change. In contrast, the RMST values of the experimental group start at 5.4 months when the threshold is equal to 0 and increase until 9.6 months when the threshold that indicates the change in the hazard ratio function is equal to 12 months. The rational for this behaviour is that if the threshold is higher, the time through the study in which $S_1(t) > S_0(t)$ is also higher which makes the average survival time of the experimental group increases. Hence, it is straightforward to interpret and understand why when the threshold is equal to 12 months, the ratio between RMST is equal to 0.91, and how as the threshold increases, the ratio between average survival times will decrease. In Table \ref{table2} we also provide the difference between the RMST of each group, where we see that, in line with the type of non-proportional hazards (i.e, crossing hazards), these difference go from negative values to positive values as the threshold increases.

\begin{table}[h]
\centering
\caption{RMST of each treatment group in a crossing hazards setting for values of the threshold that indicates the change in the hazard ratio function ranging from 0 to 12 months.}
\vspace{0.25cm}
\begin{tabular}{|c|c|c|c|c|c|}
\hline
\makecell{Threshold (months)} & 0    & 3    & 6    & 9   & 12   \\ \hline
\makecell{RMST in control group}& 7.4 & 7.5 & 7.6 & 7.7 & 7.8 \\ \hline
\makecell{RMST in experimental group}& 5.4 & 6.9 & 8.0 & 8.9 & 9.6 \\ \hline
\makecell{Difference between RMST} & -2.1 & -0.6 & 0.4 & 1.2 & 1.8 \\ \hline
\end{tabular}
\label{table2}
\end{table}

\subsubsection{Decreasing treatment effects}

\begin{figure}[h]
  \centering
  \caption{Power (A) and treatment effect (B) estimation using the standard log-rank test, weighted log-rank test and RMST in a decreasing effects setting.}
  \vspace{0.25cm}
  \includegraphics[scale=0.55]{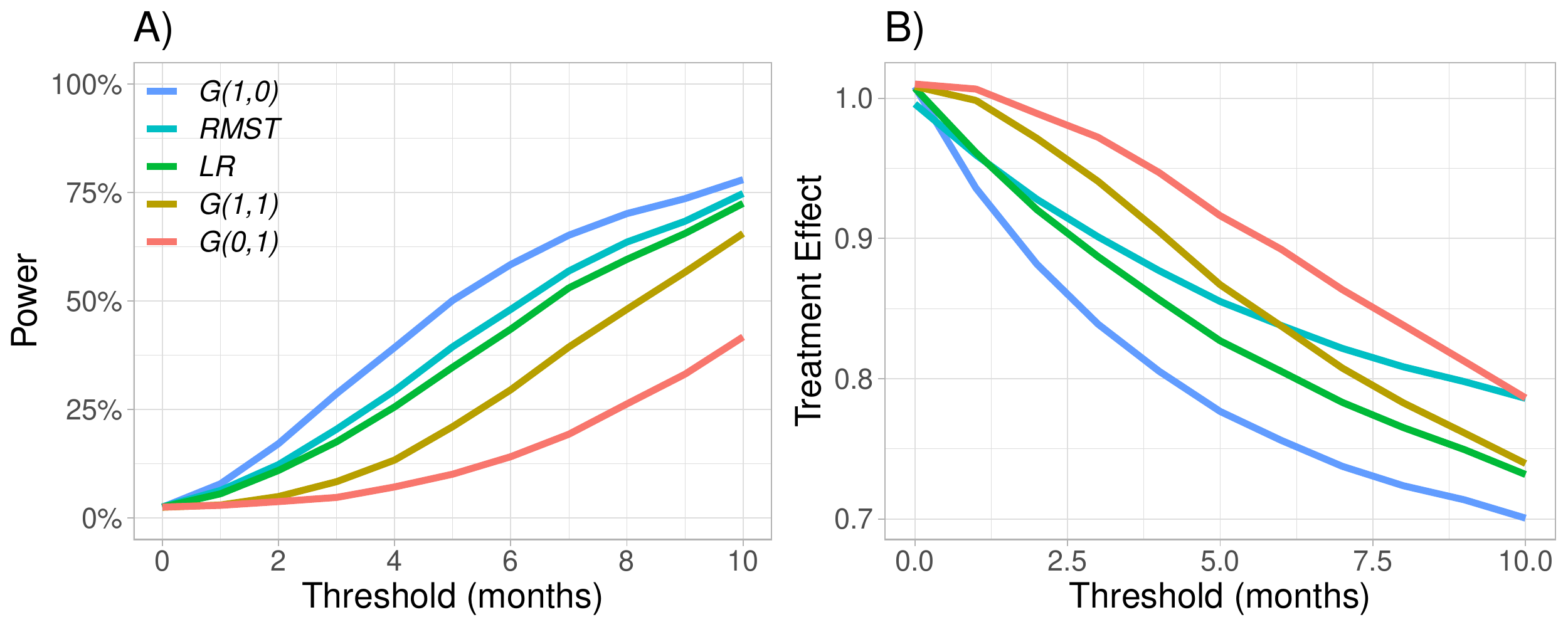}
  \label{figure7}
\end{figure}

In Figure \ref{figure7} we present the empirical power and the estimated treatment effects of the proposed scenario under decreasing effects with a threshold that indicates the change in the hazard ratio function ranging from 0 to 10 month. In the context of this type of non-proportional hazards pattern, this means that from the beginning of the study until a time threshold $S_1(t) > S_0(t)$, and after the threshold $S_1(t) = S_0(t)$ . In terms of power, we observe a general increase as the threshold that indicates the change in the hazard ratio function increases. 

This behaviour is consistent since when the threshold is equal to 0, $S_0(t) = S_1(t)$ throughout the entire and hence the probability of rejecting $H_0$ is $\alpha = 0.025$. Then, as the the threshold increases, the time throughout the study in which $S_1(t) > S_0(t)$ is also higher, and hence it is more likely to reject $H_0$, which causes a power increase. As expected, the weighted log-rank tests $G_{(1,0)}$ and $G_{(0,1)}$ are the tests that achieve highest and lowest power overall, respectively. Also, the standard log-rank test and the test based on the RMST achieve very similar power values regardless the threshold that indicates the change in the hazard ratio function.

In terms of the quantification of the treatment effect we observe that with a threshold of 0 months (i.e., $S_1(t) = S_0(t)$), all treatment effect estimations presented in in Figure \ref{figure7}B are equal to 1. Then, as the threshold increases, the estimated treatment effects decreases favoring the experimental treatment group. This behavior is consistent with the nature of this type of non-proportional hazards since the lower the threshold the earlier the full treatment effect disappears and we start to observe that $S_1(t) = S_0(t)$. The hazard ratio represented in Figure \ref{figure7}B with the ``LR'' line, has a treatment effect equal to 1 when the threshold is equal to 0 months, and equal to 0.74 when the threshold is equal to 10 months. However, under non-proportional hazards (i.e., threshold $> 0$) this would be an average hazard ratio which does not a clear clinical interpretation under non-proportional hazards.

The weighted hazard ratio represented in Figure \ref{figure7}B with the ``$G_{01}$'', ``$G_{10}$'' and ``$G_{11}$'' lines also start yield a treatment effect estimation of 1 the threshold is equal to 0. However, when the threshold is high, say 10 months, we can see how $G_{01}$ and $G_{10}$ yield a treatment effect 0.8 and 0.7 respectively. The reason for these differences is that $G_{01}$ down-weights early events and $G_{10}$ down-weights late events. Nevertheless, these results do not have a straightforward clinical interpretation even if the estimated treatment effect of $G_{10}$ seems to be closer the full effect when $S_1(t) > S_0(t)$.

The ratio between RMST is represented in Figure \ref{figure7}B with the ``RMST'' line. To have a better understanding of these values, in Table \ref{table3} we present all the RMST for each treatment group and for each threshold value considered in this setting. We observe that the RMST values of the experimental group barely change and range from 10.8 and 11.1 months. In contrast, the RMST values of the control group start at 11.1 months when the threshold is equal to 0, and decrease until 8.5 months when the threshold is equal to 10 months. 

Mind that in this non-proportional hazards setting, we observe the full treatment effect until a time-threshold in which, for example, patients in the control group switch to the experimental treatment group. Therefore, the rational for this behaviour is that if the threshold is higher, the time through the study in which $S_1(t) > S_0(t)$ is also higher, which makes the average survival time of the control group decrease since these patients would switch treatment later. 

Hence, it is straightforward to interpret and understand why when the threshold is equal to 10 months, the ratio between RMST is equal to 0.78, and how as the threshold increases, the ratio between average survival times will decrease. In Table \ref{table3} we also provide the difference between the RMST of each group, where we see that, in line with the type of non-proportional hazards (i.e, decreasing effects), these differences increase as the threshold that establish when patients in the control group switch treatment increases.

\begin{table}[h]
\centering
\caption{RMST of each treatment group in a decreasing effects setting for values of the threshold that indicates the change in the hazard ratio function ranging from 0 to 10 month.}
\vspace{0.25cm}
\begin{tabular}{|c|c|c|c|c|c|c|c|c|c|c|}
\hline
\makecell{Threshold (months)} & 0 & 2 & 4 & 6 & 8 & 10 \\ \hline
\makecell{RMST in control group} & 11.1 & 10.3 & 9.7 & 9.1 & 8.8 & 8.5 \\ \hline
\makecell{RMST in experimental group} & 11.1 & 11.1 & 11.0 & 10.9 & 10.9 & 10.8 \\ \hline
Difference between RMST & 0 & 0.8 & 1.3 & 1.8 & 2.1 & 2.3 \\ \hline
\end{tabular}
\label{table3}
\end{table}

\subsection{Type-I error}
\label{sc_type1error}

\begin{figure}[h]
  \centering
  \caption{Empirical type-I error of the log-rank test, weighted log-rank test and test based on the RMST under delayed effects (A), decreasing effects (B) and crossing hazards (C).}
  \vspace{0.5cm}
  \includegraphics[scale=0.55]{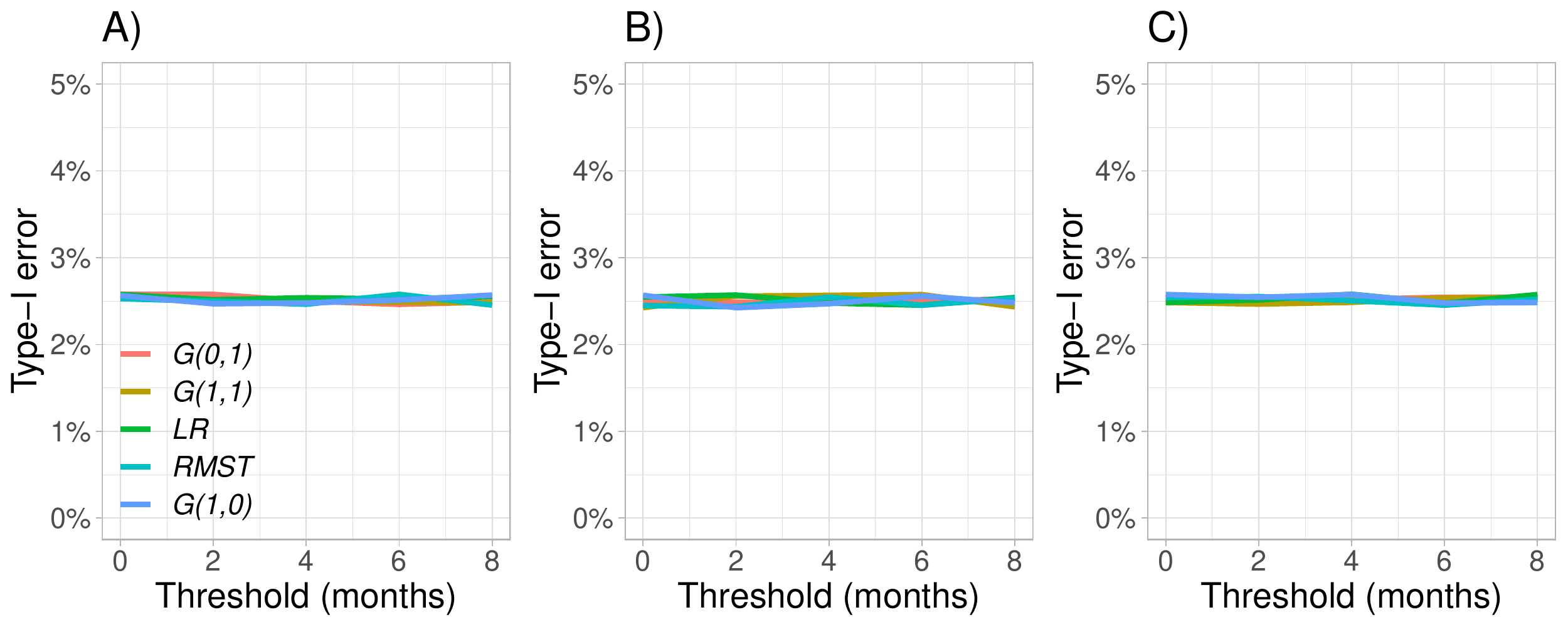}
  \label{figure8}
\end{figure}

In this section we present the empirical type-I error under the null hypothesis defined in equation \eqref{eq_h0h1} {\add for the 3 non-proportional hazards patterns presented in Figure \ref{figure4} using the simulation set-up presented in section \ref{sc_simulation_setup}, which uses a one-sided level $\alpha$ of 2.5\%. Mind that, with the simulation set-up used in this article where the control arm is the same in all non-proportional hazards patterns, we can present the empirical type-I error for all non-proportional hazards patterns in only one plot (Figure \ref{figure8}). Therefore, under $H_0$ where $S_0(t) = S_1(t)$, the survival curves in each of the non-proportional hazards patterns have the same shape. In Figure \ref{figure8}A, Figure \ref{figure8}B and Figure \ref{figure8}C we observe that type-I error rate is controlled for the standard log-rank, the weighted log-rank and the test based on the RMST either under delayed effects, decreasing effects and crossing hazards, respectively}.

In a hypothetical delayed effects setting, we may be interested in testing $H_0: \delta_0 = \delta_1 \Rightarrow S_0(t) = S_1(t)$, where $\delta$ represents the threshold that indicates the change in the hazard ratio function (i.e., delay). For example, we may be comparing an immunotherapy compound against a combination of the same immunotherapy compound and another compound that may potentially reduce the delay.

\begin{figure}[h]
  \centering
  \caption{{\add Hazard (A) and survival (B) function from the control and the experimental arm when $H_0: \delta_0 = \delta_1$.}}
  \vspace{0.5cm}
  \includegraphics[scale=0.55]{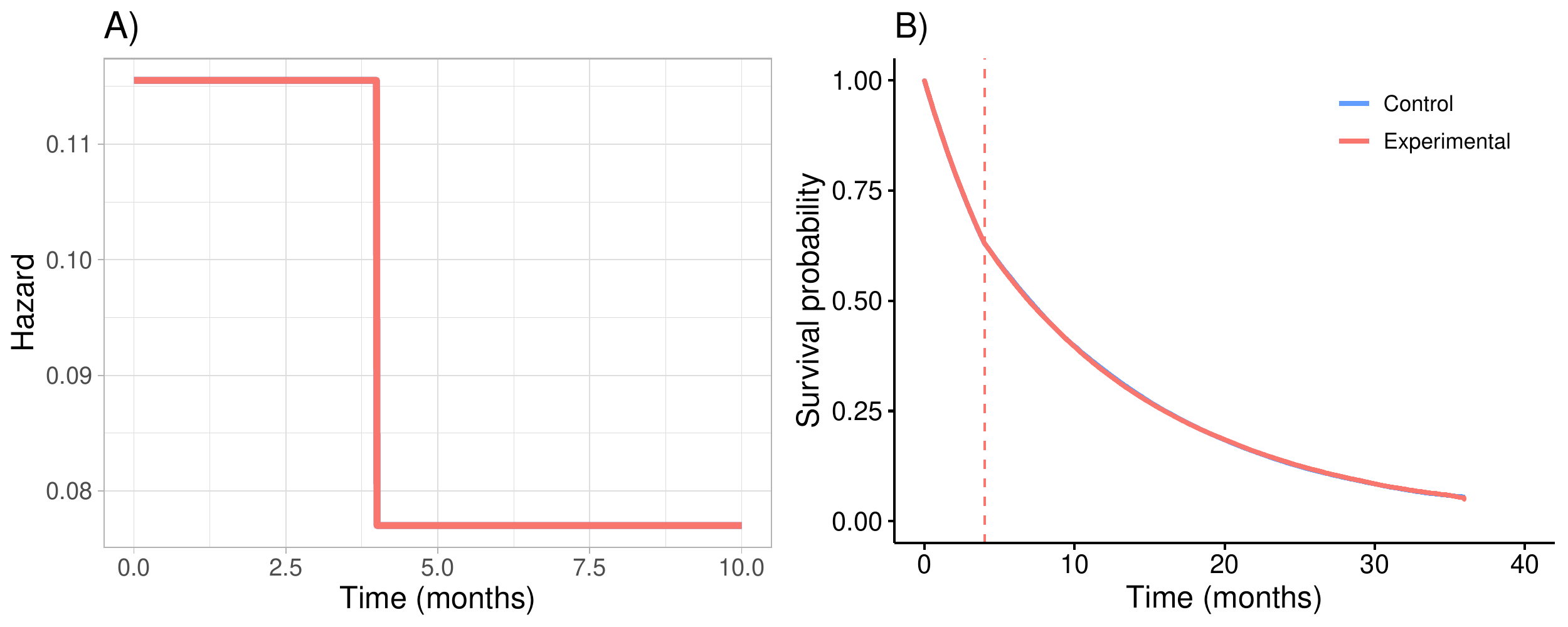}
  \label{figure9}
\end{figure}

\begin{figure}[h]
  \centering
  \caption{{\add Empirical type-I error of the log-rank test, weighted log-rank test and test based on the RMST under a delayed effects non-proportional hazards pattern with the null hypothesis $H_0: \delta_0 = \delta_1$. The type-I error is calculated using different time-points (thresholds) that represent the moment at which the hazard function of each treatment group changes.}}
  \vspace{0.5cm}
  \includegraphics[scale=0.55]{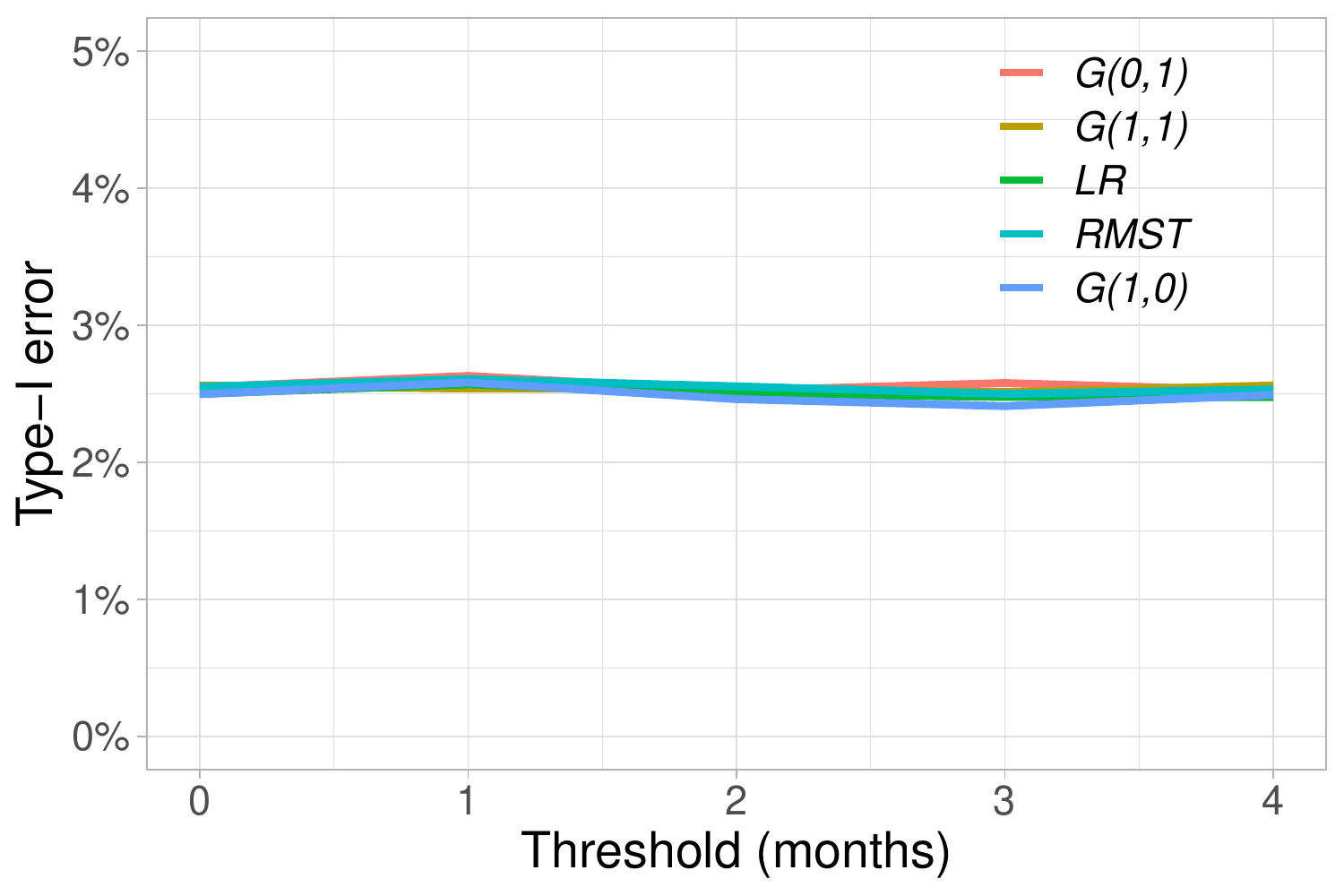}
  \label{figure10}
\end{figure}

{\add Under $H_0: \delta_0 = \delta_1$, the hazard of each treatment group is presented in Figure \ref{figure9}A using the same simulation set-up used in section \ref{sc_simulation_setup} and assuming, for illustrating purposes, a delay equal to 4 months. The corresponding survival functions are presented in Figure \ref{figure9}B. Under this other type of $H_0$ we see how both hazard functions and survival functions completely overlap. In Figure \ref{figure10} we see how, under this type of $H_0$, the type-I error is perfectly controlled when using the log-rank test, the weighted log-rank test and the test based on the RMST.

Alternative null hypotheses such us $H_0: \delta_0 \leq \delta_1 \Rightarrow S_1(t) \leq S_0(t)$ could be tested. However, a proper type-I error assessment is recommended since, as pointed out by \cite{magirr2019modestly} (Figure 5), type-I error is not controlled in a delayed effects setting when using the weighted log-rank test using the Fleming and Harrington class of weights under a null hypothesis of the type $H_0: S_1(t) \leq S_0(t)$.}

\section{Discussion}

Despite the extensive existing literature about statistical methods to deal with non-proportional hazards, the log-rank test and the hazard ratio are generally considered the gold-standard approaches even if they implicitly assume a constant treatment effect (i.e., proportional hazards) over time. However, it is well known that when treatment effect is not constant over time, the log-rank test loses power although it is a valid test. 

Under a delayed effects' type of non-proportional hazards, it is argued in the literature that we may account for the impact of the non-proportional hazards by increasing the sample size. This may be feasible with relatively small delays, but with medium to large delays the cost related to the additional sample size would be too high and hence would not be feasible. However, non-proportional hazards not only have an impact on the power; the hazard ratio becomes a biased estimate of the treatment effect and looses its clinical interpretation.

Non-proportional hazards are not rare nowadays in clinical trials, specially with the development of immunotherapy compounds, which are not typically directed to the tumor itself and aim to boost the patient's immune system. Potential positive effects with these compounds may not be observed immediately and the lag between the activation of immune cells, their proliferation, and impact on the tumor are described in the literature as a delayed treatment effect.

The weighted log-rank test together with the Fleming and Harrington class of weights (see \cite{fleming1981class}) have gained a lot of attention over the last year as they allow to down-weight early, middle or late events which can result in a power increase with respect to the standard log-rank test while controlling type-I error rate depending on the non-proportional hazards pattern and the null hypothesis under consideration (see \cite{magirr2019modestly}). Nevertheless, the choice of $\rho$ and $\gamma$ in the Fleming and Harrington class of weights requires extensive knowledge of the shape of survival curves and plays a key role that may result in loss of power of the weighted log-rank test. Therefore, due to the uncertain nature of non-proportional hazards, specification of $\rho$ and $\gamma$ is difficult. To preserve the integrity of the trial, these parameters should be pre-specified in advance. However, due to their complex interpretation, it would be interesting to explore whether, in a group sequential design setting, these parameters could be updated based on interim data since there exist methods to control the type-I error such us the conditional error function approach (see \cite{bauer2004modification, wassmer2006planning, magirr2016sample,jimenez2020response}).

In this article we have also discussed how it is possible to use a weighted version of the hazard ratio using the Fleming and Harrington class of weights. The weighted hazard ratio can be thought as an average weighted treatment effect, although it does not have a straightforward clinical interpretation plus it has inherent ethical problems as it implies that some patients’ lives are less important than others.

Another approach that has gained a lot of attention over the last years is the restricted mean survival time (RMST), a well-established, yet underutilized measure that can be interpreted as the average survival time up to a pre-specified, clinically important time point. It is equivalent to the area under the Kaplan-Meier curve from the beginning of the study until a cutting point. In this article we have discussed how it is not as powerful has the weighted log-rank test although it achieves an almost identical performance compared to the standard log-rank test, which is nowadays gold-standard approach. However, the RMST most important benefit over the hazard ratio and the weighted hazard ratio is that it is always interpretable, even under non-proportional hazards. 

One of the most common critiques against the RMST is that the statistical significance of the results depends on the cutting time,  which  is  constrained  to  the  duration  of  the follow-up and the censoring. However, the log-rank test is a sequence of hypergeometric tests that requires both the number of patients at risk and number of events to be greater than zero (see \cite{fleming2011counting}). This implies that the log-rank test (and by extension the weighted log-rank test) depend on a time-window. Moreover, the maximum time-window allowed by the RMST is actually wider than the one for the log-rank test, which means that the RMST uses more data than the log-rank and weighted log-rank tests (see \cite{huang2020estimating}).

However, using a relative risk measure (i.e., the hazard ratio, the weighted hazard ratio or the ratio based on the RMST) has its own limitation as it does not provide absolute benefit of the treatment effect. It is well known that relative risk measures have the advantage of being stable across populations with different baseline risks. This is particularly useful when, for example, we aim to combine the results of different trials in a meta-analysis. However, they have the major disadvantage that do not reflect the baseline risk of the individuals with respect to the outcome that is being measured. In other words, relative risk measures such us the hazard ratio do not take into consideration the individuals’ risk of achieving the outcome of interest without the administration of a drug and therefore do not really tell how much benefit the individual would obtain from its use.

{\add We may consider the robust milestone (or landmark) analysis and the RMST difference as simple and clinically interpretable alternatives to the relative risk measures mentioned above. The robust milestone analysis would have the difficulty of selecting the milestone, although several options exist in the literature to protect the study against a poor choice of the milestone (see e.g.,  \cite{magirr2020non}).}

One topic we would like to explore further in the future is the use of the weighted log-rank test in a delayed effects setting with small sample sizes. The weighted log-rank test gives higher weights to late survival times. This may make the test susceptible to small sample sizes as the latter part of the survival curve is estimated based on the smaller sample sizes (i.e., very small number of “at-risk” groups later part of the study follow-up). In this case, we wonder whether the RMST may have some advantage over the weighted log-rank test in this situation in terms of power.

\section*{Data availability}

The \texttt{R} code used in this article is available at \texttt{\href{https://github.com/jjimenezm1989/Quantifying-treatment-differences-in-confirmatory-trials-under-non-proportional-hazards}{https://github.com/jjimenezm1989}}.

\section*{Acknowledgements}

I am grateful to the Associate Editor and one anonymous referee for their constructive comments on earlier versions of this manuscript.

\section*{Disclaimer}

The views and opinions expressed in this article are those of the author and do not necessarily reflect the official policy or position of Novartis Pharma A.G.

\bibliographystyle{tfs}
\bibliography{references}

\begin{thebibliography}{10}
\providecommand{\MR}{\relax\unskip\space MR }
\providecommand{\url}[1]{\normalfont{#1}}
\providecommand{\urlprefix}{Available at }

\bibitem{bauer2004modification}
P. Bauer and M. Posch, \emph{Modification of the sample size and the schedule
  of interim analyses in survival trials based on data inspections, by h.
  sch{\"a}fer and h.-h. m{\"u}ller, statistics in medicine 2001; 20:
  3741-3751.}, Statistics in Medicine 23 (2004), p. 1333.

\bibitem{brown1984choice}
M. Brown, \emph{On the choice of variance for the log rank test}, Biometrika 71
  (1984), pp. 65--74.

\bibitem{fleming1981class}
T.R. Fleming and D.P. Harrington, \emph{A class of hypothesis tests for one and
  two sample censored survival data}, Communications in Statistics-Theory and
  Methods 10 (1981), pp. 763--794.

\bibitem{fleming2011counting}
T.R. Fleming and D.P. Harrington, \emph{Counting processes and survival
  analysis}, Vol. 169, John Wiley \& Sons, 2011.

\bibitem{freidlin2019methods}
B. Freidlin and E.L. Korn, \emph{Methods for accommodating nonproportional
  hazards in clinical trials: ready for the primary analysis?}, Journal of
  Clinical Oncology 37 (2019), p. 3455.

\bibitem{harrington1982class}
D.P. Harrington and T.R. Fleming, \emph{A class of rank test procedures for
  censored survival data}, Biometrika 69 (1982), pp. 553--566.

\bibitem{huang2018comparison}
B. Huang and P.F. Kuan, \emph{Comparison of the restricted mean survival time
  with the hazard ratio in superiority trials with a time-to-event end point},
  Pharmaceutical statistics 17 (2018), pp. 202--213.

\bibitem{huang2020estimating}
B. Huang, L.J. Wei, and E.B. Ludmir, \emph{Estimating treatment effect as the
  primary analysis in a comparative study: Moving beyond p value}, Journal of
  clinical oncology: official journal of the American Society of Clinical
  Oncology 38 (2020), pp. 2001--2002.

\bibitem{irwin1949standard}
J. Irwin, \emph{The standard error of an estimate of expectation of life, with
  special reference to expectation of tumourless life in experiments with
  mice}, Epidemiology \& Infection 47 (1949), pp. 188--189.

\bibitem{jimenez2020modified}
J.L. Jim{\'e}nez, J. Niewczas, A. Bore, and C.F. Burman, \emph{A modified
  weighted log-rank test for confirmatory trials with a high proportion of
  treatment switching}, arXiv preprint arXiv:2005.09213  (2020).

\bibitem{jimenez2019properties}
J.L. Jim{\'e}nez, V. Stalbovskaya, and B. Jones, \emph{Properties of the
  weighted log-rank test in the design of confirmatory studies with delayed
  effects}, Pharmaceutical statistics 18 (2019), pp. 287--303.

\bibitem{jimenez2020response}
J.L. Jim{\'e}nez, V. Stalbovskaya, and B. Jones, \emph{Response to comments on"
  properties of the weighted log-rank test in the design of confirmatory
  studies with delayed effects" by jos{\'e} l. jim{\'e}nez, viktoriya
  stalbovskaya and byron jones, pharmaceutical statistics, 2019; 18: 287-303},
  Pharmaceutical statistics  (2020).

\bibitem{karrison1987restricted}
T. Karrison, \emph{Restricted mean life with adjustment for covariates},
  Journal of the American Statistical Association 82 (1987), pp. 1169--1176.

\bibitem{latimer2016treatment}
N.R. Latimer, C. Henshall, U. Siebert, and H. Bell, \emph{Treatment switching:
  statistical and decision-making challenges and approaches}, International
  journal of technology assessment in health care 32 (2016), pp. 160--166.

\bibitem{lee1996some}
J.W. Lee, \emph{Some versatile tests based on the simultaneous use of weighted
  log-rank statistics}, Biometrics  (1996), pp. 721--725.

\bibitem{lin2017estimation}
R.S. Lin and L.F. Le{\'o}n, \emph{Estimation of treatment effects in weighted
  log-rank tests}, Contemporary clinical trials communications 8 (2017), pp.
  147--155.

\bibitem{magirr2020non}
D. Magirr, \emph{Non-proportional hazards in immuno-oncology: is an old
  perspective needed?}, arXiv preprint arXiv:2007.04767  (2020).

\bibitem{magirr2019modestly}
D. Magirr and C.F. Burman, \emph{Modestly weighted logrank tests}, Statistics
  in medicine 38 (2019), pp. 3782--3790.

\bibitem{magirr2016sample}
D. Magirr, T. Jaki, F. Koenig, and M. Posch, \emph{Sample size reassessment and
  hypothesis testing in adaptive survival trials}, PloS one 11 (2016).

\bibitem{pak2017interpretability}
K. Pak, H. Uno, D.H. Kim, L. Tian, R.C. Kane, M. Takeuchi, H. Fu, B. Claggett,
  and L.J. Wei, \emph{Interpretability of cancer clinical trial results using
  restricted mean survival time as an alternative to the hazard ratio}, JAMA
  oncology 3 (2017), pp. 1692--1696.

\bibitem{royston2011use}
P. Royston and M.K. Parmar, \emph{The use of restricted mean survival time to
  estimate the treatment effect in randomized clinical trials when the
  proportional hazards assumption is in doubt}, Statistics in medicine 30
  (2011), pp. 2409--2421.

\bibitem{royston2013restricted}
P. Royston and M.K. Parmar, \emph{Restricted mean survival time: an alternative
  to the hazard ratio for the design and analysis of randomized trials with a
  time-to-event outcome}, BMC medical research methodology 13 (2013), p. 152.

\bibitem{schemper1992cox}
M. Schemper, \emph{Cox analysis of survival data with non-proportional hazard
  functions}, Journal of the Royal Statistical Society: Series D (The
  Statistician) 41 (1992), pp. 455--465.

\bibitem{schemper2009estimation}
M. Schemper, S. Wakounig, and G. Heinze, \emph{The estimation of average hazard
  ratios by weighted cox regression}, Statistics in medicine 28 (2009), pp.
  2473--2489.

\bibitem{schoenfeld1981asymptotic}
D. Schoenfeld, \emph{The asymptotic properties of nonparametric tests for
  comparing survival distributions}, Biometrika 68 (1981), pp. 316--319.

\bibitem{tian2020empirical}
L. Tian, H. Jin, H. Uno, Y. Lu, B. Huang, K.M. Anderson, and L. Wei, \emph{On
  the empirical choice of the time window for restricted mean survival time},
  Biometrics  (2020).

\bibitem{tian2014predicting}
L. Tian, L. Zhao, and L. Wei, \emph{Predicting the restricted mean event time
  with the subject's baseline covariates in survival analysis}, Biostatistics
  15 (2014), pp. 222--233.

\bibitem{trinquart2016comparison}
L. Trinquart, J. Jacot, S.C. Conner, and R. Porcher, \emph{Comparison of
  treatment effects measured by the hazard ratio and by the ratio of restricted
  mean survival times in oncology randomized controlled trials}, Journal of
  Clinical Oncology 34 (2016), pp. 1813--1819.

\bibitem{uno2014moving}
H. Uno, B. Claggett, L. Tian, E. Inoue, P. Gallo, T. Miyata, D. Schrag, M.
  Takeuchi, Y. Uyama, L. Zhao, \emph{et~al.}, \emph{Moving beyond the hazard
  ratio in quantifying the between-group difference in survival analysis},
  Journal of clinical Oncology 32 (2014), p. 2380.

\bibitem{uno2015alternatives}
H. Uno, J. Wittes, H. Fu, S.D. Solomon, B. Claggett, L. Tian, T. Cai, M.A.
  Pfeffer, S.R. Evans, and L.J. Wei, \emph{Alternatives to hazard ratios for
  comparing the efficacy or safety of therapies in noninferiority studies},
  Annals of internal medicine 163 (2015), pp. 127--134.

\bibitem{wassmer2006planning}
G. Wassmer, \emph{Planning and analyzing adaptive group sequential survival
  trials}, Biometrical Journal: Journal of Mathematical Methods in Biosciences
  48 (2006), pp. 714--729.

\bibitem{xu2017designing}
Z. Xu, B. Zhen, Y. Park, and B. Zhu, \emph{Designing therapeutic cancer vaccine
  trials with delayed treatment effect}, Statistics in medicine 36 (2017), pp.
  592--605.

\bibitem{zhao2016restricted}
L. Zhao, B. Claggett, L. Tian, H. Uno, M.A. Pfeffer, S.D. Solomon, L. Trippa,
  and L. Wei, \emph{On the restricted mean survival time curve in survival
  analysis}, Biometrics 72 (2016), pp. 215--221.

\bibitem{zucker1998restricted}
D.M. Zucker, \emph{Restricted mean life with covariates: modification and
  extension of a useful survival analysis method}, Journal of the American
  Statistical Association 93 (1998), pp. 702--709.

\end{thebibliography}

\clearpage

\end{document}